\documentclass{iau}

\usepackage{graphicx}
\usepackage{multirow}
\usepackage{natbib}
\graphicspath{{./}{images/}}

\def\b#1,{{\bf #1,}}
\def\afterjournal{ \b}

\def\apjs {{\it Astrophys.\ J.\ Suppl.}\afterjournal}
\def\apjl {{\it Astrophys.\ J.\ Lett.}\afterjournal}

\def\nrdc#1.{1986, in ``Light on Dark Matter'', ed. F.P. Israel,
  (Reidel, Dordrecht), p #1.}
\def\calc#1.{1986, in ``Late Stages of Stellar Evolution'', eds. S. Kwok and
  S.R. Pottasch (Reidel, Dordrecht), p #1.}
\def\camc#1.{1987, in ``Galaxy'', eds. ?. ? (Reidel, Dordrecht), p #1.}
\def\torc#1.{1988, in ``Mass of the Galaxy'', ed. M. Fich,
  (Toronto University Press), p #1.}

\newcommand{\kms}{\,km s$^{-1}$}
\newcommand{\micron}{\,$\mu$m}%
\newcommand{\arcsec}{$^{\prime\prime}$}

\begin{document}

\lefttitle{Ritter et al.}
\righttitle{From an amateur PN candidate to the Rosetta Stone of SN Iax research}

\journaltitle{Planetary Nebulae: a Universal Toolbox in the Era of Precision Astrophysics}
\jnlPage{1}{7}
\jnlDoiYr{2023}
\doival{10.1017/xxxxx}
\volno{384}

\aopheadtitle{Proceedings IAU Symposium}
\editors{O. De Marco, A. Zijlstra, R. Szczerba, eds.}
 

\aopheadtitle{Proceedings IAU Symposium}

\title{From an amateur PN candidate to the Rosetta Stone of SN Iax research}

\author{A. Ritter$^1$, Q. A. Parker$^1$, F. Lykou$^2$, A. A. Zijlstra$^3$, M. A. Guerrero$^4$, P. Le D\^u$^5$}
\affiliation{$^1$Laboratory for Space Research, FoS, The University of Hong Kong, Hong Kong (S.A.R.)\\
$^2$ Konkoly Observatory, Research Centre for Astronomy and Earth
Sciences, Budapest, Hungary\\
$^3$ The University of Manchester, Manchester, U.K.\\
$^4$ Instituto de Astrof\'isica de Andaluc\'ia (IAA-CSIC), Granada, Spain\\
$^5$ Kermerrien Observatory, Porspoder, France}







\begin{abstract}
On August 25$^{\rm th}$ 2013 Dana Patchick from the “Deep Sky Hunters” (DSH) amateur astronomer group discovered a diffuse nebulosity in the Wide-field Infrared Survey Explorer (WISE) mid-IR image archive that had no optical counterpart but appeared similar to many Planetary Nebulae (PNe) in WISE. As his 30th discovery he named it Pa 30 and it was added to the HASH PN database as a new PN candidate. Little did he know how important his discovery would become. 10 years later this object is the only known bound remnant of a violent double WD merger accompanied by a rare Type Iax SN, observed and recorded by the ancient Chinese and Japanese in 1181 AD. This makes Pa 30 and its central star IRAS 00500+6713 (WD J005311) the only SN Iax remnant in our Galaxy, the only known bound remnant of any SN, and based on the central star's spectrum the only Wolf-Rayet star known that neither has a massive progenitor nor is the central star of a Planetary Nebula. We cover this story and our key role in it.
\end{abstract}


\maketitle

\section{Introduction}

While the number of known Supernovae (SNe) of the type Iax is relatively small, they constitute a significant fraction of the total number of SNe Ia, $\sim 5\%–30\%$ \citep{
Foley2013,White2015}. Yet they are probably the least understood subclass of type Ia SNe. The distribution of host-galaxy morphologies for SNe Iax is strongly skewed toward star-forming, late-type host galaxies \citep[e.g.][and references therein]{White2015}, which may indicate massive star progenitors for this class \citep[e.g.][]{Valenti2009}.
However, the maximum-light spectra of SNe Iax show clear evidence for C/O burning, providing a strong link between SNe Iax and White Dwarf (WD) progenitors \citep{Foley2010}.
While the exact mechanisms leading to a Type~Iax SN are still not fully understood, they are believed to arise from either the failed detonation of a CO WD accreting material from a Helium donor star \citep[single degenerate scenario, e.g. ][]{Kromer2015} or from a CO WD merging with a heavier Oxygen-Neon (ONe) WD \citep[double degenerate scenario, ][]{kashyap2018} where the accretion disk itself deflagrates. All SNe Iax observed so far are in distant galaxies and they quickly fade below detectability, even with the biggest telescopes. Using X-ray spectroscopy, \citet{Zhou2021} found high Mn/Fe and Ni/Fe ratios in the Galactic SNR Sgr A East, interpreting it as the remnant from an Iax event involving a CO WD. However, \citet{Zhang2023} have casted doubt on that interpretation given its low X-ray luminosity. Identifying a Galactic SNR of the type Iax is of utmost importance for understanding this type of SN.
Here we will chronologically line out the path from an amateur PN candidate (\ref{sec:amateur}) to a WD merger remnant (\ref{sec:gvaramadse}) to a SNR (\ref{sec:snr}) to the historical SN 1181 AD (\ref{sec:SN1181}) to a bound remnant of a SN Iax in our Galactic neighborhood (\ref{sec:Iax}). Our latest results are presented in (\ref{sec:lykou}), and the conclusions in (\ref{sec:conclusions}).

\section{Pa 30 as an amateur PN candidate}
\label{sec:amateur}
When Dana Patchick \citep{kronberger2016} from our affiliated "Deep Sky Hunters`` (DSH) amateur group discovered a diffuse nebulosity (Fig.~\ref{fig:colour}) in the Wide-field Infrared Survey Explorer (\emph{WISE}) mid-IR image archive on August 25$^{\rm th}$ 2013, little did he realise what he had found. As Patchick's 30th PN candidate it was originally included into the HASH database {\footnote{HASH the consolidated 'gold-standard' PN data repository and catalogue: 
\url{http://www.hashpn.space}} \citep{hash} as Pa\,30 as it possessed certain PN-like characteristics \citep[but see][]{Parker2022}. The \emph{WISE}  W3 (11\micron) band  shows a  disk-like nebula, whereas the dominant shape in the  W4 (22\micron) band is donut-like inside a much fainter halo. Narrow-band [O~{\sc iii}] imaging observations obtained by the DSH group on the 2.1m KPNO telescope in September 2013 (see Fig.~\ref{fig:oiii_sii}) indicate very faint circular emission with hints of radial structure confirmed in the [S~{\sc ii}] images in \citep{Fesen2023}.
\begin{figure*}
	\centering
	\includegraphics[width=0.8\textwidth]{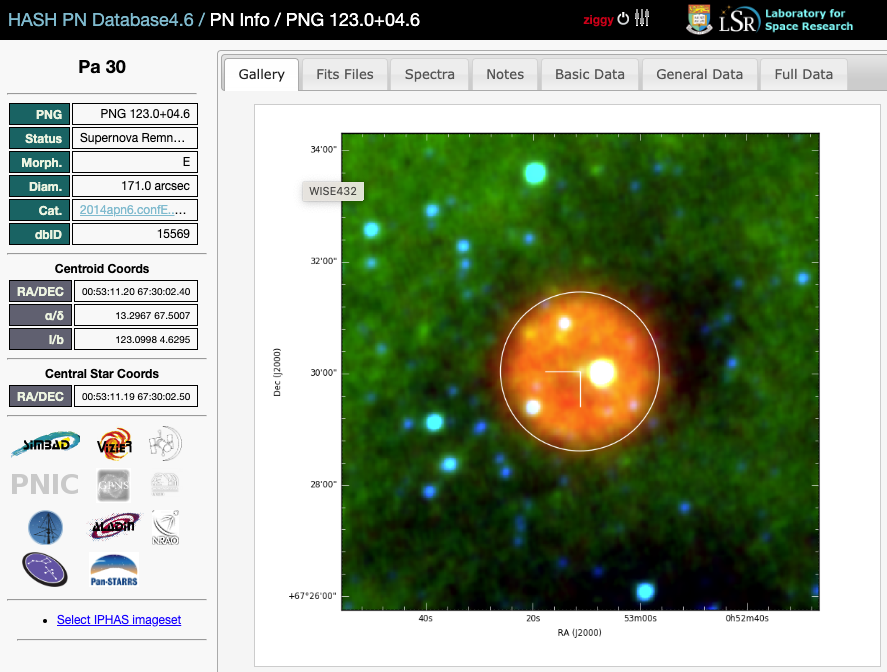}
	\caption{False colour image of Pa\,30 and Parker's star in the HASH database. \textit{Blue} stands for \emph{WISE/W2}, \textit{green} for \emph{WISE/W3}, and \textit{red} for \emph{WISE/W4}.}
 \label{fig:colour}
\end{figure*}

\begin{figure*}
	\centering
	\includegraphics[height=0.35\textwidth]{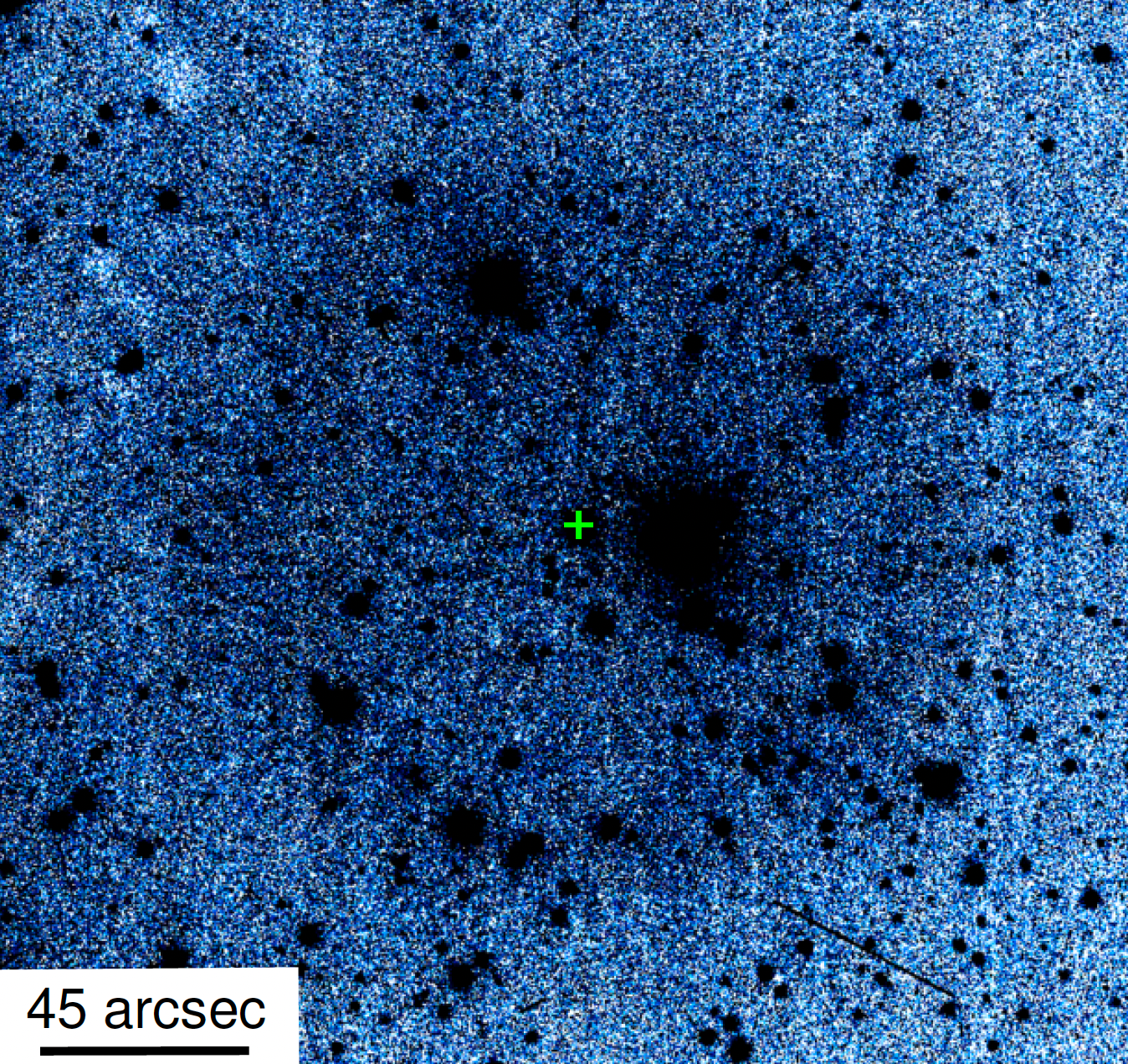}\includegraphics[height=0.35\textwidth]{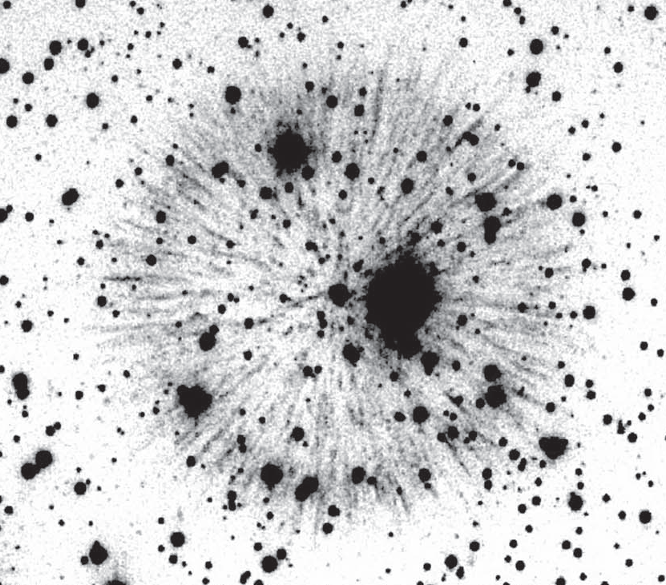}
	\caption{Left: The 2.1-m KPNO [O~{\sc iii}] image, which we have stacked and rebinned from individual frames to enhance the low surface brightness, diffuse shell though with hints of radial structure that are seen clearly in the [S~{\sc ii}] image of \citep[][Right]{Fesen2023}. The green cross in the center of the left image marks the location of the CS. }
 \label{fig:oiii_sii}
\end{figure*}


On October 15$^{\rm th}$ 2014, the DSH team used the SparsePak integral field unit (IFU) on the 3.5-m WIYN telescope at KPNO  to observe Pa\,30 as a PN candidate. They did not see the expected PN emission lines and did not extract the central star's spectrum.

We obtained deep optical spectroscopy of both the star and the nebula on July 8$^{\rm th}$ 2016 using long slit observations with the OSIRIS instrument of the 10-m GranTeCan (GTC) telescope.
The spectra of the central star are shown in Fig.~\ref{fig:spectra}.\\
\begin{figure*}
	\centering
	\includegraphics[width=0.7\textwidth]{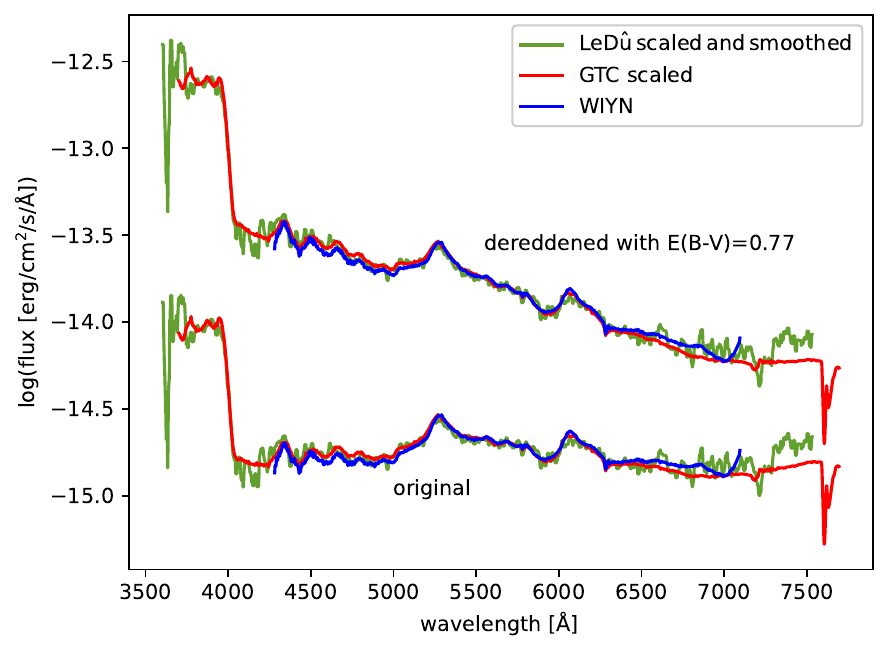}
	\caption{Stellar spectra from SparsePak/WIYN, OSIRIS/GTC, and a small amateur telescope (P.~Le D\^u) smoothed and scaled to the same flux level. The dereddened spectra are also shown for comparison \citep[using interstellar extinction from ][]{Ritter2021}. The 
	major discontinuity bluewards of 4,000\,\AA\ is seen in the OSIRIS and amateur spectra but the SparsePak spectrum did not go below 4,200\,\AA. The OSIRIS spectrum had 
	the slit slightly off-center to avoid a nearby bright star. To correct for the resulting slight flux underestimation after flux calibration the spectrum was scaled to match the reliable SparsePak/WIYN flux level. From \citet{Lykou2023}
	}
	\label{fig:spectra}
\end{figure*}
The spectrum of the CS is dominated by broad emission lines of highly ionised Oxygen (O VI to O VIII) and Carbon (C IV to C VI) lines. The measured wind velocities of 16,000 \kms\ (0.05 c) are unprecedented. On careful scrutiny both sets of our spectroscopic observations revealed two sets of faint optical emission lines species in the nebula: the [S {\sc ii}] 6716 and 6731\AA\ doublet, and the [Ar {\sc iii}] 7136\AA\ line. The well resolved [S {\sc ii}] doublet in particular shows two sets of lines due to extreme velocity expansion of the nebula gas, with radial velocities of up to $v_{\rm rad}\approx$1,100$\pm$100 \kms\ (Fig.~\ref{fig:vrad}). The observed velocity structure shows sharp variations across the field of view and declines to systemic velocity at $\approx$100\arcsec. We also extracted the star's GTC spectrum after its remarkable features were first brought to our attention from a spectrum taken by Pascal Le D\^u, an amateur collaborator of the corresponding author. We recognised its extreme and unique nature  \citep[e.g. see][]{gvaramadze2019,Lykou2023}, believing we were the first group to do so. We thereafter referred to IRAS 00500+6713 as ``Parker's star".\\

\begin{figure*}
	\centering
		\includegraphics[width=\textwidth]{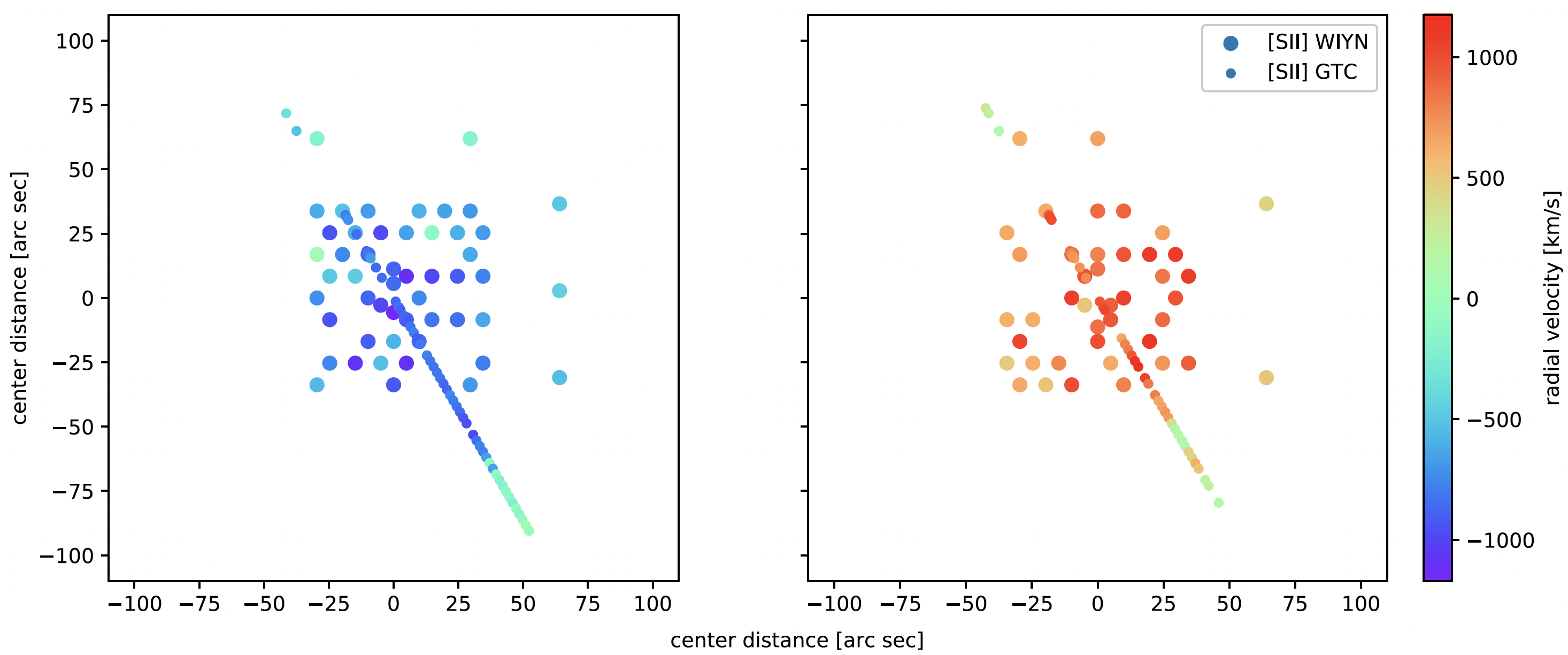}
	\caption{Position-velocity diagram of the [S {\sc ii}] doublet lines (6,716/6,731 \AA) from WIYN/SparsePak, overplotted with GTC/OSIRIS along the slit, color-coded to the radial velocity for each measured fibre/CCD column. North is up, East is left. The left/right plots show the near/far side of the nebula. 
The images show sharp velocity variations
where the two sets of [S {\sc ii}] doublet lines are also much brighter than for the rest of the nebula, indicating strong shock excitation both for the near as well as the far side.
}
	\label{fig:vrad}
\end{figure*}

\section{Pa 30's Central Star as a WD merger remnant}
\label{sec:gvaramadse}
While our Nature Astronomy paper on Pa\,30 and its CS was in second review, \citet{gvaramadze2019} published their own Nature paper on the same source. They proposed that the CS was from a double-degenerate CO + ONe WD merger. They identified the CS as an oxygen-rich Wolf Rayet (WO-type) star. The star's spectrum was modelled with the Potsdam Wolf Rayet (PoWR) code and they derived a stellar temperature at the base of the 16,000 \kms\ wind of $211,000^{+40,000}_{-23,000}$ K. They determined that the chemical composition is dominated by oxygen and carbon with mass fractions of $0.8\pm0.1$ and $0.2\pm0.1$, respectively. No Hydrogen or Helium lines were detected as expected for WD merger remnants \citep[e.g.][]{Schwab2016}. As driver for the fierce stellar wind they proposed a strong magnetic field of the order of $10^8$~G.

\section{Pa\,30 as a SNR}
\label{sec:snr}
The same group then published X-ray observations of Pa\,30 and its CS \citep{oskinova2020}. In their X-ray nebula spectra they detected carbon-burning ashes, indicating that Pa\,30 is a supernova remnant (SNR), possibly a type Iax, and that the double WD merger was accompanied by an episode of carbon burning. From SNR scaling relations they estimated an age of $\sim$1,000 years. Given the \emph{GAIA} DR2 distance of 3.1~kpc they estimated that the SN would have been visible for 2 weeks and so could have been missed by the ancient astronomers. They did not check the historical records.

\section{Pa 30 and the historical SN 1181 AD}
\label{sec:SN1181}
The only SN of the last millennium that did not have a firmly identified counterpart on the sky was the historical SN~1181~AD. 
It was described in 3 ancient Chinese and 1 Japanese record that it looked like Saturn and lasted for 185 days. Averaging the best positional estimates from all 4 records places SN~1181 AD in the vicinity of Pa\,30. The only other SNR possibly linked to SN~1181 is the pulsar wind nebula G130.7+3.1 \citep[3C58 hereafter, e.g.][]{stephenson1999,Kothes2013}. However, a precise estimate of the expansion age of this nebula of 7,000 years, based on radio observations over 20~years \citet{2006ApJ...645.1180B}, and the spin-down age of the pulsar \citep[5,400 years,][]{2004AdSpR..33..456C} have put the association in serious doubt, although not completely excluding it \citep{Kothes2013}. \citet{Ritter2021} showed that Pa\,30 and its CS are almost certainly the true remnants of the 1181~AD event. This was later supported by \citet{Schaefer2023}, who analysed the historical records in more detail (Fig.~\ref{fig:schaefer_map}), and \citet{Fesen2023}, who calculated an age estimate of 844$\pm$55 years, in excellent agreement with the current 842 years age of SN 1181.

\begin{figure*}
	\centering
		\includegraphics[width=0.5\textwidth]{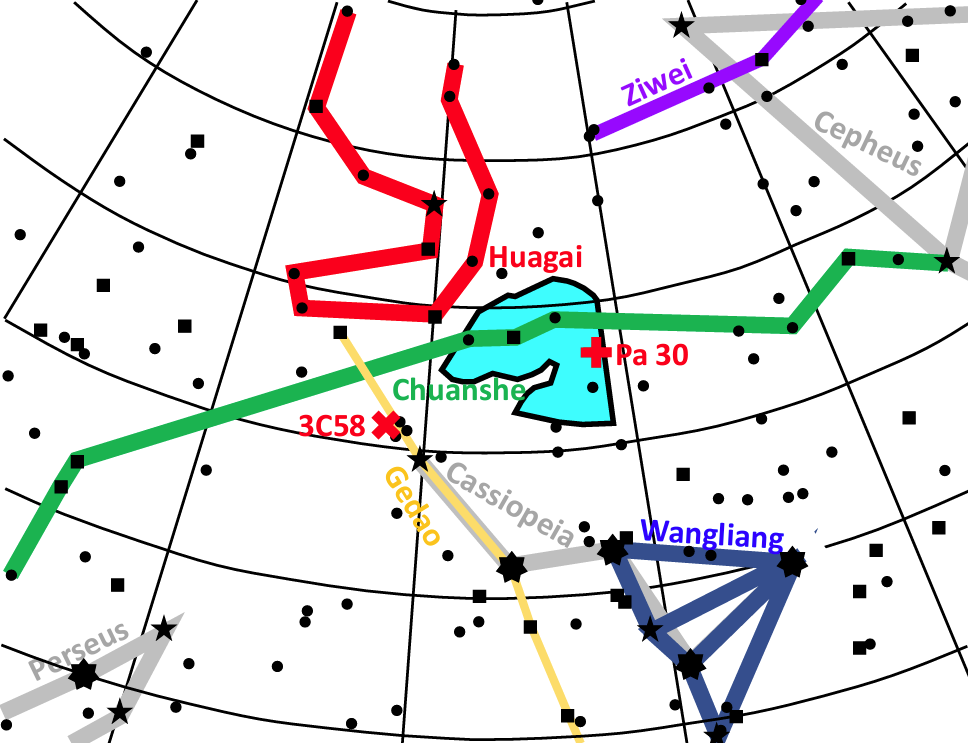}
	\caption{The final map of the most likely position of SN~1181~AD (cyan area) from the detailed interpretation of historical records by \citet{Schaefer2023}. Pa\,30 is inside this region, 3C58 is not.
}
	\label{fig:schaefer_map}
\end{figure*}

\section{Pa 30 as type Iax SNR}
\label{sec:Iax}
With the new \emph{GAIA} eDR3 distance of 2.3 kpc \citep{bailerjones2021}, the maximum absolute brightness \citep[$M_V \sim -14$ to $-12.5$,][]{Ritter2021} and duration of naked-eye visibility \citep[185 days,][]{Hsi1957}, together with the expansion velocity \citep[1,100$\pm$100 \kms,][]{Ritter2021} and inferred kinematical age \citep[$990^{+280}_{-220}$ years,][]{Ritter2021} of the surrounding nebula, are all consistent with a type Iax SN in 1181 AD.

\section{Ruling out a strong magnetic field and kicked remnant star}
\label{sec:lykou}
In \citet{Lykou2023} we analysed our high signal-to-noise spectroscopic data of both star and nebula which were obtained with the OSIRIS instrument on the 10-m Gran Telescopio Canarias (GTC) telescope in 2016, as well as archival ultraviolet spectra from the Space Telescope Imaging Spectrograph (STIS) on the Hubble Space Telescope (HST) under program GO15864 (P.I. G. Graefener; epoch: 4 November 2020). We modeled the stellar wind with the NLTE code {\sc cmfgen} adapted to WC and WO stars \citep{hillier2012b} and an additional high temperature ($\sim$4MK) gas component using the {\sc cloudy} code \citep[ver. 17.03][]{cloudy}. The best fit ranges of the individual parameters are shown in Table~\ref{tab:models}, together with a comparison to the parameters stated by \citet{gvaramadze2019}. We provided an upper limit for a magnetic field of $B<2.5\,$MG, far below the suggestions by \citet{gvaramadze2019}. We also rule out the possibility that the CS received a substantial kick, and suggest that the stellar remnant has dimmed by $\sim$0.5 magnitudes over the last 100 years.

\begin{table*}[htbp]
\centering
	\caption{Range of stellar parameters for {\sc cmfgen} models. The last two columns show the values from \citet{gvaramadze2019} and from the hot gas model. From \citet{Lykou2023} 
	}
 \label{tab:models}
 \resizebox{\textwidth}{!}{%
	\begin{tabular}{lccccc}
	\hline
	Parameter & Explored range & Best fit range & \citet{gvaramadze2019} & 
	Hot gas model \\
	\hline
$L_*$ ($\rm L_{\odot}$) & 10,000 -- 200,000  & 30,000 -- 60,000 &  
$39,810^{+20,144}_{-10,970}$ & \\
$T_*$ (K) & 145,000 -- 580,000 & 200,000 -- 280,000   &  237,000 & 211,000$^{+40,000}_{-23,000}$ & 
\\
$\dot{M}$ ($\rm M_{\odot}$/yr) & 7.5$\times$ 10$^{-7}$ -- 2.5$\times$ 10$^{-6}$ & $\leq$ 4$\times$ 
10$^{-6}$ & $3.5(\pm0.6)\times 10^{-6}$ & \\
$R_*$ ($\rm R_{\odot}$) & 0.04 -- 0.22 & $\leq 0.2$ & 0.15$\pm$0.04 & \\
v$_{\infty}$ (kms) & -- & $\sim$15,000 & 16,000$\pm$1,000 & \\
\hline
	Mass fractions & & & \\
	\hline
$X_{\rm H}$ &  --  & -- & -- & -- \\
$X_{\rm He}$ &  0.017 -- 0.135 & $<0.4$ & $<0.1$ & $\leq 0.44 $\\
$X_{\rm C}$ &  0.0 -- 0.261 & $\leq 0.26$ & $0.2\pm0.1$ & 0.13 \\
$X_{\rm O}$ &  0.414 -- 0.697 & $\leq 0.7 $ &  $0.8\pm0.1$ & 0.39  \\
$X_{\rm Ne}$ & 0.011 -- 0.501 & $\leq 0.5 $ & 0.01 & 0.04 \\
\hline
	\end{tabular}}
\end{table*}

\section{Conclusions}
\label{sec:conclusions}
WD~J005311 (Parker's star) and Pa\,30 are the result of a violent double-degenerate merger (ONe/CO or CO/CO WDs) accompanied by a Type~Iax SN. It remains the only Iax SNR known in our Galaxy and the only one that can be studied in detail. The CS is also the only bound SNR and the only known WR star which has neither a massive progenitor star nor is the CS of a PN. It has an unprecedented wind speed of 16,000 \kms (0.05 c) and is the hottest star known ($>$220,000 K). The CS was not kicked as predicted by most WD merger models. The lack of a companion star \citep[see ][]{Schaefer2023} strongly supports the double degenerate merger scenario for SN~Iax.
\bibliography{bibliography}{}
\bibliographystyle{aasjournal}

\end{document}